\documentclass[reprint,
showkeys,
amsmath,
amssymb,
aip,
jcp,
]{revtex4-1}

\usepackage{graphicx}
\usepackage{bm}
\usepackage[page]{appendix}
\usepackage{amsmath}
\usepackage{scalerel}
\usepackage{subfigure}
\usepackage{color}
\usepackage[colorlinks=true,allcolors=blue,urlcolor=blue]{hyperref}

\begin{document}

\title{Generalized Langevin equation for a tagged monomer in a Gaussian semiflexible polymer}

\author{Xavier Durang}
\affiliation{Department of Physics, Pohang University of Science and Technology (POSTECH), Pohang 37673, Republic of Korea}
\email{durangx@gmail.com}
\author{Jae-Hyung Jeon}
\affiliation{Department of Physics, Pohang University of Science and Technology (POSTECH), Pohang 37673, Republic of Korea}
\affiliation{Asia Pacific Center for Theoretical Physics (APCTP), Pohang 37673, Republic of Korea}
\email{jeonjh@gmail.com}
\date{\today}

\begin{abstract}
In this study, we present a comprehensive analysis of the motion of a tagged monomer within a Gaussian semiflexible polymer model. We carefully derived the generalized Langevin Equation (GLE) that governs the motion of a tagged central monomer. This derivation involves integrating out all the other degrees of freedom within the polymer chain, thereby yielding an effective description of the viscoelastic motion of the tagged monomer. A critical component of our analysis is the memory kernel that appears in the GLE. By examining this kernel, we characterized the impact of bending rigidity on the non-Markovian diffusion dynamics of the tagged monomer. Furthermore, we calculated the mean-squared displacement of the tagged monomer using the derived GLE. Our results not only show remarkable agreement with previously known results in certain limiting cases but also provide dynamic features over the entire timescale.
\end{abstract}

\maketitle


\section{\label{sec:intro}Introduction}
Semiflexible biopolymers have recently regained interest with an increased number of studies investigating their physical properties. Semiflexible polymers, such as dsDNA, F-actin, or microtubules, are typically long-chain molecules that exhibit different conformations depending on their flexibility. This flexibility is characterized by a persistence length comparable to their contour length, approximately 50 nm \cite{Bust1994}, 16 $\mu$m\cite{Ott1993}, and 5 mm \cite{Gitt1993} for dsDNA, F-actin, and microtubules respectively. It influences the structural and dynamical properties of polymers and therefore plays a crucial role in biological processes such as protein assembly, DNA dynamics, enzymatic catalysis, and filament organization \cite{Gara2020, Mark1995, Pete2010, Shus2004, Rich2019, Vlie2020, Clae2006, Wink2020}. 
Among the models employed for studying the behaviors of semiflexible polymers, the worm-like chain (WLC) model, introduced by Kratky and Porod,  stands out \cite{Krat1949}. This model represents polymer conformation by an inextensible and differentiable curve with bending flexibility and has been used to simulate biomolecules such as dsDNA and microtubules \cite{Kremer1990,Mark1995,Wilh1996,Bath2008,Kosl2014,Petr2006,Hinc2009,Mara2018}. However, exact analytical derivations are not achievable, due to its non-linearity, necessitating simulations or approximations for most dynamical properties \cite{Klei2006, Hsu2010, Likh2012, Mara2018}.

Thus, from a theoretical perspective, alternative models of semiflexible polymers have been proposed. A model, similar to the WLC but constraining the average contour length instead of the bond length, was introduced and was analytically trackable \cite{Harr1966, Hear1967}. The dynamics of linear semiflexible chains was also examined using linear response methods \cite{Bixo1978} or path integral formulations \cite{Free1972, Klei2006}, allowing calculation of most conformational statistics at equilibrium. The dynamics of another bead-spring model\cite{Bark2012, Bark2014}, with a Hamiltonian approximated in the first order by the Hessian, was also studied and it showed similar behavior as the WLC.

The model that we study is the Gaussian semiflexible polymer model \cite{Winkler1994a, Winkler1994b, Harnau1995}. In this model, linear chains of beads are connected by harmonic springs and subjected to bending rigidity similar to the WLC but constraining the mean-square bond length instead of the individual bond length. This model captures the essential features of semiflexible polymers, such as the transition from rod-like to coil-like behavior with the bending rigidity parameter $\kappa$. However, in the limit of high stiffness, it exhibits a distinct behavior from that of the WLC due to its Gaussian character. This model has found extensive application in diverse contexts, including polymer solutions, networks, dendritic polymers, polymer translocations, and polymer interactions \cite{Wink2003, Furs2012, MacKintosh1995, Dolg2009, Dolg2009a,Dolg2010, Kroy1997, Eman2007, Lopez2018}. Moreover, other studies have generalized it to incorporate nonlinear effects, branched structures, ring topologies, active dynamics, and shear flow \cite{Ever2004, Grest1986, Harn1996, Halv2011, Ghosh2014, Eise2016, Schro2005, Wink2020}. Additionally, it has been compared with experimental data and other theoretical models \cite{Mark1995}.

The Generalized Langevin Equation (GLE) has attracted significant attention for understanding diffusion dynamics within these polymer systems. It is represented as $\int_0^t K(t-t')\dot{x}(t')dt' = \Xi(t)$ where $\Xi(t)$ denotes a noise process whose autocovariance is intricately linked to the memory kernel through the Fluctuation-Dissipation Theorem (FDT), specifically expressed as $K(t-t') = k_B T\langle\Xi(t)\Xi(t')\rangle$. The memory kernel, denoted by $K(t)$, captures the viscoelastic relaxation of the surrounding environment. At timescales relevant to the collective behavior of many-body interactions, this memory kernel often follows a power-law behavior, characterized by $K(t) \sim t^{-\alpha}$, where $0 < \alpha < 1$. For instance, it has been analytically demonstrated using various methods~\cite{Liza2010, Maes2013, Panja2010, Vand2015} that a single monomer within a flexible polymer, or Rouse polymer, is governed by $K(t) \sim t^{-1/2}$. Extensions of these findings have been made to incorporate additive active noise, either acting on each beads~\cite{Vand2015, Vand2017a, Vand2017b}, or affecting a single bead~\cite{Joo2020}. However, the generalization of these results to Gaussian semiflexible polymers remains unexplored, despite the knowledge that in such polymers, the memory kernel displays a distinctive behavior, $K(t) \sim t^{-3/4}$ in Refs.~\cite{Gran1997, Casp2000, Han2023}. Recent studies have addressed the dynamics of a tagged monomer in a polymer chain~\cite{Dolg2011, Bull2011, Furs2013, Bark2014, Mart2019, Wink2010, Teje2023}. However, these works directly compute some macroscopic quantities and do not rely on the GLE formalism while other studies~\cite{Colm2015, Han2023, Joo2023} employ the GLE as a starting equation. Here, we want to address the analytical derivation of the GLE of a tagged monomer in a finite semiflexible chain and then use this GLE to understand the tagged monomer's dynamics over the entire timescale.

In this study, our primary objective is to provide an analytical derivation of the Generalized Langevin Equation (GLE) from the bead-spring model and provide a comprehensive characterization of non-Markovian effects across various regimes. To accomplish this, we extend the works previously done on the Rouse polymer \cite{Doi1988, Vand2015, Vand2017a} to the case of the Gaussian semiflexible polymer. The structure of this paper is as follows: In Section II, we introduce the Gaussian model of semiflexible polymers within the framework of a discrete representation, where beads are connected by springs that tend to align with their neighboring beads. We then provide an explicit derivation of the GLE. In Section III, we analyze the memory kernel, derive its behavior in each of the regimes, and discuss the limiting case of a stiff polymer. Moving on to Section IV, we utilize the derived GLE to compute the mean square displacement and discuss the influence of the bending rigidity in this context.

\section{\label{sec:semiflexible}Generalized Langevin Equation}
In this section, after defining the Gaussian semiflexible polymer, we demonstrate that the evolution equation of a tagged monomer of a finite semiflexible polymer is a GLE. More specifically, we will detail the procedure for the case of the central monomer, because for sufficiently long chains, the dynamics of a tagged monomer should be the same as the dynamics of the central one.

\subsection{Definition of the semiflexible Gaussian polymer}
\begin{figure}[ht]
\centering
\includegraphics[width=0.98\linewidth]{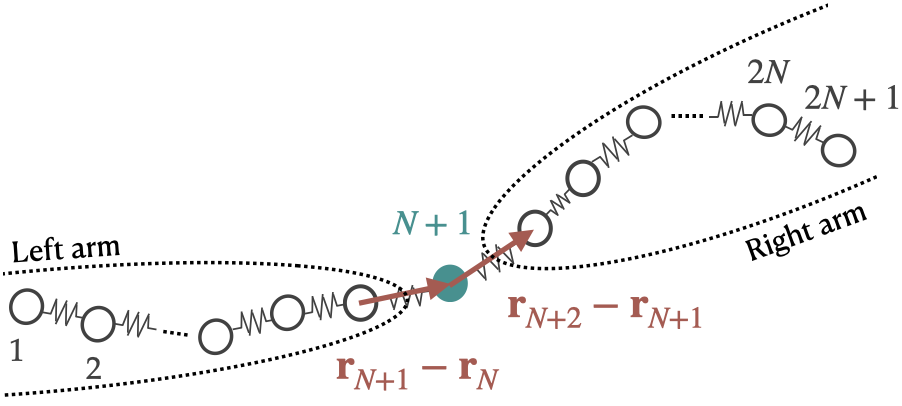}
\caption{Schematic of a semiflexible Gaussian chain of $2N+1$ beads. The central bead $N+1$, whose position is given as $\mathbf{r} = \mathbf{r}_{N+1}$, connects the left arm ${\mathbf{R}^L}=(\mathbf{r}_0, \mathbf{r}_1, ..., \mathbf{r}_N)$ to the right arm ${\mathbf{R}^R}=(\mathbf{r}_{N+2}, \mathbf{r}_{N+3}, ..., \mathbf{r}_{2N+1})$.}
	\label{fig:Fig1}
\end{figure}

The Hamiltonian of the semiflexible Gaussian chain model \cite{Wink1994,Harn1995} is a quadratic function of the bead positions, which enables the exact solution of the chain dynamics in terms of the appropriate normal modes. The linear semiflexible chain consists of $2N+1$ identical beads, linked by $2N$ harmonic springs. The position vector of the center of each bead at time $t$ is $\mathbf{r}_i(t)$, where $i=1,\dots,2N+1$, and the bond vector between two adjacent beads is $\mathbf{q}_i=\mathbf{r}_{i+1}-\mathbf{r}_{i}$, where $i=1,\dots,2N$. The potential energy of the bonds in the model can be expressed as 
\begin{equation}\label{eq:potsemiflex}
	U = \frac{k}{2}\left[\sum_{i=1}^N\lambda_i\mathbf{q}_i^2 - \sum_{i=1}^{N-1}\mu_i\mathbf{q}_i\mathbf{q}_{i+1}\right].
\end{equation}
 where the first term represents the elastic energy of the harmonic springs, and the second term $-\mu_i\mathbf{q}_i\mathbf{q}_{i+1}$ is a penalty energy for misalignment. Here, $k$ is the spring constant $k=\frac{3k_BT}{b^2}$. The parameters $ \lambda_i,\,\mu_i $ in \eqref{eq:potsemiflex} are Lagrange multipliers that enforce the constraints applied to the bonds. There are two constraints 
\begin{equation}\label{eq:requirements}
\begin{split}
	&\langle\mathbf{q}_i^2\rangle = b^2\qquad i = 1,...,2N+1\\
	&\langle\mathbf{q}_i\mathbf{q}_{i+1}\rangle = \kappa b^2\qquad i = 1,...,2N.
\end{split}
\end{equation}
The first constraint imposes $b^2$ as the mean square bond length, whereas the second constraint defines the correlations between two successive bonds, where $\kappa$ is a dimensionless parameter such that $\kappa = \langle \cos\theta_i\rangle$, where $\theta_i$ is the angle between consecutive bond vectors $\mathbf{q}_{i}$ and $\mathbf{q}_{i+1}$, see Fig.~(\ref{fig:Fig1}). The parameter $\kappa$ regulates the bending rigidity of the chain, satisfying $0\leq \kappa <1$.  We list two limiting cases: when $\kappa =0$, we recover the flexible Rouse chain. When $\kappa \rightarrow1$, the polymer resembles a rigid rod, although its contour length fluctuates.
The Lagrange multipliers were determined using the maximum entropy principle \cite{Wink1994,Wink2003,Dolg2009}
\begin{equation}\label{eq:lagrangemultipliers}
\begin{split}
\lambda_i=\frac{1}{2}\frac{1+\kappa^2}{1-\kappa^2},\quad i=2,...2N\\
\mu_i =\frac{\kappa}{1-\kappa^2},\quad i=1,...N-1,
\end{split}
\end{equation}
and $\lambda_1=\lambda_{2N+1}=\frac{1}{2(1-\kappa^2)}$.

For later discussions, we introduce two key parameters associated with the polymer: the contour length denoted as $L=Nb$, and the persistence length, designated as $L_p=b/(1-\kappa)$. By defining the dimensionless parameter $l = L/L_p$, we can readily distinguish the two limiting cases: (i) When $l \gg 1$, the polymer exhibits flexibility. In this scenario, the contour length is much longer than the persistence length, signifying that the polymer is characterized by its ability to bend: It is the flexible Rouse chain. (ii) Conversely, when $l \ll 1$, the polymer exhibits stiffness. In this case, the contour length is much shorter than the persistence length, indicating that the polymer is less inclined to bend and is instead characterized by a more rigid and constrained structure (Fig.~\ref{fig:Fig1}).

\subsection{Derivation of the GLE for the central bead}
The dynamics of each bead of the polymer is governed by  the Langevin equation
\begin{equation}\label{eq:Langevinposition}
\gamma \frac{\partial \mathbf{r}_n}{\partial t}=-\mathbf{\nabla} U + \mathbf{\xi}_n
\end{equation}
where the potential $U$ is given by Eq.~(\ref{eq:potsemiflex}) and $\xi_n$ represents the thermal noise acting on the $n^{th}$ bead. These thermal forces follow a Gaussian distribution with a zero average. These forces are uncorrelated between  different beads and their correlation satisfies the fluctuation-dissipation relation 
\begin{equation}
\langle \xi_n(t) \xi_m(t') \rangle = 6\gamma k_BT\delta(t-t')\delta_{n,m}.
\end{equation}
To construct an effective generalized Langevin equation for the central bead $\mathbf{r} = \mathbf{r}_{N+1}$, we have to integrate out all the degrees of freedom of the other beads. For that matter, it is convenient to separate the potential energy as follows 
\begin{equation}
  U = U_{\rm ch} + U_{\rm cb}
\end{equation}
where  $U_{\rm cb} $ regroups all the terms of the potential energy Eq.~(\ref{eq:potsemiflex}) which involves the central bead and its expression is 
\begin{eqnarray}\label{eq:cbpot}
U_{\rm cb} &=& \frac{k}{2}\left[ (2\lambda +\mu){\mathbf{r}}^2 -(2\lambda +2 \mu) (\mathbf{r}_N +\mathbf{r}_{N+2}){\mathbf{r}} \right. \\ \nonumber 
& & \left. + \mu(\mathbf{r}_{N-1}+\mathbf{r}_{N+3}){\mathbf{r}} \right].
\end{eqnarray}
Thus, $U_{\rm ch} $ regroups all other terms (meaning all terms of the potential energy in Eq.~(\ref{eq:potsemiflex}) which does not involve the central bead). We rewrite it as 
\begin{equation}\label{eq:Uch}
U_{\rm ch} = \frac{k}{2}\left(\mathbf{R}^L H^L {\mathbf{R}^L}^T +\mathbf{R}^R H^R {\mathbf{R}^R}^T \right)
\end{equation}
which is the potential of the two chains, left and right respectively denoted by an index $^L$ and $^R$, surrounding the central bead as in Fig.~\ref{fig:Fig1}. ${\mathbf{R}^{L,R}}$ are the vectors of the position coordinates for each bead in the left ${\mathbf{R}^L}=(\mathbf{r}_0, \mathbf{r}_1, ..., \mathbf{r}_N)$ and in the right arm ${\mathbf{R}^R}=(\mathbf{r}_{N+2}, \mathbf{r}_{N+3}, ..., \mathbf{r}_{2N+1})$. The matrices $H^{L,R}$ are the coupling matrices whose expressions are given in Appendix A.
As the coupling matrices $H^{L,R}$ are Hermitian and normal, diagonal matrices $D^{L,R}$ must exist such that $D^{L,R}={A^{L,R}}^{-1}H^{L,R}A^{L,R}$ where $A^{L,R}$ are unitary matrices. Therefore, solving the matrix equation $A^{L,R}D^{L,R} =H^{L,R} A^{L,R}$ will yield the eigenvalues $\lambda_p$, found on the diagonal of the matrices $D^{L,R}$, and $A^{L,R}$ will be composed of the eigenvectors. However, to simplify the problem, we use the Rouse normal modes as eigenvectors
\begin{equation}\label{eq:rousemode}
\mathbf{r}_n = \frac{2}{\sqrt{2N+1}}\sum_{p=1}^{N} c_n^p\mathbf{X}_p^L
\end{equation}
with $c_n^p={\rm cos} \left(\frac{2p-1}{2N+1}\left(n-\frac{1}{2}\right)\pi \right) $.
These modes diagonalize the potential energy for $\mu =0$ and are known to asymptotically diagonalize it in the semi-flexible case for a large number of beads $N$ \cite{Shil2002}. With this simplification, from $A^{L,R}D^{L,R} =H^{L,R} A^{L,R}$, we obtain the following eigenvalues 
\begin{equation}
\lambda_p = 4{\rm sin}^2\left( \frac{2p-1}{2N+1}\frac{\pi}{2}\right)\left[\lambda-\mu + 2\mu {\rm sin}^2\left( \frac{2p-1}{2N+1}\frac{\pi}{2}\right) \right].
\end{equation}

To obtain a closed equation for the central bead, the next step is to obtain an equation of motion for the normal modes. It reads
\begin{eqnarray}
\gamma \frac{d{\mathbf X_p}^{(L)}}{dt}&=&\frac{2\gamma}{\sqrt{2N+1}}\sum_{n=1}^N c_n^p\frac{d {\mathbf r_n}}{dt}
\\ \nonumber
&=&\frac{2}{\sqrt{2N+1}}\sum_{n=1}^{N} c_n^p \xi_n +\frac{k}{2}\left[\left(4\lambda+2\mu \right) c_n^p {\mathbf r_{n}} \right.
\\ 
& & + \left(2\lambda +2\mu\right) c_n^p\left({\mathbf r_{n-1}}+{\mathbf r_{n+1}}\right)
\\ \nonumber
& & \left. -\mu c_n^p\left({\mathbf r_{n-2}}+{\mathbf r_{n+2}}\right)\right].
\end{eqnarray}
This expression can be further developed using the diagonalization condition $A^{L,R}D^{L,R} =H^{L,R} A^{L,R}$ and neglecting all terms involving the boundaries. We get
\begin{eqnarray}\label{eq:setmodes}
\gamma \frac{d{\mathbf X_p}^{(L)}}{dt}&=&-k\lambda_p{\mathbf X_p}^{(L)} + \frac{(2\lambda+2\mu)k}{\sqrt{2N+1}}\sum_{n=1}^N \left[ (2\lambda+2\mu)c_n^p{\mathbf r} \right.
 \nonumber  \\
& & \left.-\mu \left(c_{N-1}^p{\mathbf r} + c_N^p({\mathbf r_{N+2}}+{\mathbf r_N})\right)\right] + \hat{\xi}_p^{(L)}
\end{eqnarray}
where $\hat{\xi}_p^{(L)} = \frac{2}{\sqrt{2N+1}}\sum_{n=1}^{N} c_n^p \xi_n$. If $\mu =0$, the equation closes and we can recover the known result valid for the Rouse chain \cite{Vand2017a,Maes2013, Panja2010}. If $\mu\neq0$, the set of equations~(\ref{eq:setmodes}) does not close because of the term involving ${\mathbf r_{N+2}}+{\mathbf r_N}$. In order to close it, we simply approximate ${\mathbf r_{N+2}}+{\mathbf r_N}$ by its intermediate value ${\mathbf r_{N+2}}+{\mathbf r_N} \simeq 2 {\mathbf r_{N+1}} = 2{\mathbf r}$. This is the central approximation of the derivation. For $\mu \rightarrow 0 $, this approximation is known to provide the exact result as this term vanishes because of the prefactor $\mu$. In the limit of a rod-like polymer $\kappa \rightarrow 1$, this approximation becomes exact. However, for an intermediate value, the quality of this approximation is not well controlled therefore we can only expect to obtain a qualitative result.
After using the approximation, the closed equation is
\begin{eqnarray}
\gamma \frac{d{\mathbf X_p}^{(L)}}{dt}&=& \hat{\xi}_p^{(L)} -k\lambda_p{\mathbf X_p}^{(L)} 
\\ \nonumber& & - \frac{k(-1)^p a_p}{\sqrt{2N+1}}\left[2(\lambda-\mu)+4\mu a'_p\right]{\mathbf r}
\end{eqnarray} 
where $a_p = {\rm sin} \left(\frac{2p-1}{2N+1}{\pi} \right) $ and $a'_p = {\rm sin} \left(\frac{2p-1}{2N+1}\frac{\pi}{2} \right) $. The solution of the normal modes immediately reads
\begin{eqnarray}\label{eq:solmode}
{\mathbf X_p}^{(L)}(t)&=&{\mathbf X_p}^{(L)}(0) e^{-\frac{t}{\tau_p}}+\frac{1}{\gamma}\int_0^t d\tau \hat{\xi}_p^{(L)}(\tau) e^{\frac{t-\tau}{\tau_p}} \\ \nonumber 
&-&\frac{k(-1)^pa_p}{\gamma\sqrt{2N+1}}\left(2(\lambda-\mu)+4\mu a'_p\right) \int_0^td\tau {\mathbf r(\tau)} e^{\frac{\tau-t}{\tau_p}}
\end{eqnarray}
where $\tau_p = \frac{\gamma}{k\lambda_p}$. Up to this point, we have established the time evolution equations for the normal modes of the left and right chains connected to the central bead. To derive the time evolution of the central bead, as expressed in Eq.~(\ref{eq:Langevinposition}), and considering the energy potential given by Eq.~(\ref{eq:cbpot}), we use the expression for the normal modes provided in Eq.~(\ref{eq:rousemode}). Setting $\xi(t) = \xi_{N+1}(t)$, we obtain

\begin{eqnarray} \nonumber
\gamma \frac{d{\mathbf r}}{dt}&=&-2k(2\lambda+\mu){\mathbf r}+k\left[\left(2\lambda+2\mu\right)\left({\mathbf r_N}+{\mathbf r_{N+2}}\right) \right.\\ \nonumber
& &\left. -\mu\left({\mathbf r_{N-1}}+{\mathbf r_{N+3}}\right)\right] + \xi(t)\\
&=&-2k(2\lambda+\mu){\mathbf r} + \xi(t)
\\ \nonumber
& &-\sum_{p=1}^{N} \frac{2k(-1)^pa_p}{\sqrt{2N+1}}\left[(\lambda-\mu)+2\mu {a'}_p^2\right] \left({\mathbf X}_p^{(L)} + {\mathbf X}_p^{(R)}\right).
\end{eqnarray}
Inserting the solution of the normal modes Eq.~(\ref{eq:solmode}), the equation for the evolution of the central bead can be written as
\begin{eqnarray}\label{eq:GLE}
\gamma \dot{\mathbf r} = -\int_0^t d\tau \; K(t-\tau) \dot{\mathbf r}(\tau) + \Xi(t)+ \xi(t)
\end{eqnarray}
with the memory kernel 
\begin{eqnarray} \label{eq:kernel}
K(t) = \frac{2k}{2N+1}\sum_{p=1}^N& \frac{a_p^2}{{a'_p}^2}\left(2(\lambda-\mu)+4\mu {a'_p}^2\right) 
\\ \nonumber
&  \times e^{-\frac{4kt}{\gamma}{a'_p}^2\left(2(\lambda-\mu)+4\mu {a'_p}^2\right)}.
\end{eqnarray}
The noise term $\xi(t)$ represents the initial thermal noise acting on the central bead and satisfies the fluctuation-dissipation relation $\langle \xi(t) \xi(s) \rangle = 6k_BT\delta(t-s)$. The additional noise term $\Xi(t)$ appearing in this equation arises from the averaging process applied to the remaining degrees of freedom associated with the other beads. This term can be characterized as an effective noise, and its expression is given by:\begin{eqnarray}\label{eq:effectivenoise} 
\nonumber
&\Xi(t) = -\frac{2k}{\sqrt{2N+1}}\sum_{p=1}^N (-1)^pa_p\left(2(\lambda-\mu)+4\mu{a'_p}^2\right)e^{-\frac{t}{\tau_p}} \\
&\times \left[{\mathbf X}_p^{(L)} +{\mathbf X}_p^{(R)} +\frac{2\tau_p}{\gamma}\frac{2k}{\sqrt{2N+1}}\left(2(\lambda-\mu)+4\mu{a'_p}^2\right){\mathbf r}(0) \right. \nonumber \\
&\left.+\frac{1}{\gamma}\int_0^t d\tau \left(\hat{\xi}_p^{(L)}+\hat{\xi}_p^{(R)}\right)e^{\frac{\tau}{\tau_p}} \right]
\end{eqnarray}
and following the same lines as in \cite{Vand2017a}, one can show that it satisfies a fluctuation-dissipation relation with the memory kernel such that $ \langle \Xi(t) \Xi(s) \rangle = 3k_BTK(|t-s|)$ at large times.
Eqs.~(\ref{eq:GLE}--\ref{eq:effectivenoise}) constitute the main result of this paper. Because the effective noise Eq.~(\ref{eq:effectivenoise}) satisfies the fluctuation-dissipation theorem at large times, the behavior of the noise correlation will be the same as that of the kernel. Therefore, the only quantity that needs to be studied is the kernel which we analyze in the following section.

\section{Analysis of the kernel}
\subsection{Generic behavior of the kernel}
In this section, we focus on the general expression of the kernel $K(t)$ defined in Eq.~(\ref{eq:kernel}). To facilitate our analysis, we decompose the kernel into two distinct contributions
 \begin{eqnarray} \label{eq:semikernel1}
K_1(t) = \frac{4(\lambda-\mu)  k}{2N+1}\sum_{p=1}^N& \frac{a_p^2}{{a'_p}^2}e^{-\frac{8kt}{\gamma}{a'_p}^2\left((\lambda-\mu)+2\mu {a'_p}^2\right)} \\ \label{eq:semikernel2}
K_2(t) = \frac{8\mu k}{2N+1}\sum_{p=1}^N& a_p^2 e^{-\frac{8kt}{\gamma}{a'_p}^2\left((\lambda-\mu)+2\mu {a'_p}^2\right)}.
\end{eqnarray}
\begin{figure}[t]
	\centering
	\includegraphics[width=\columnwidth]{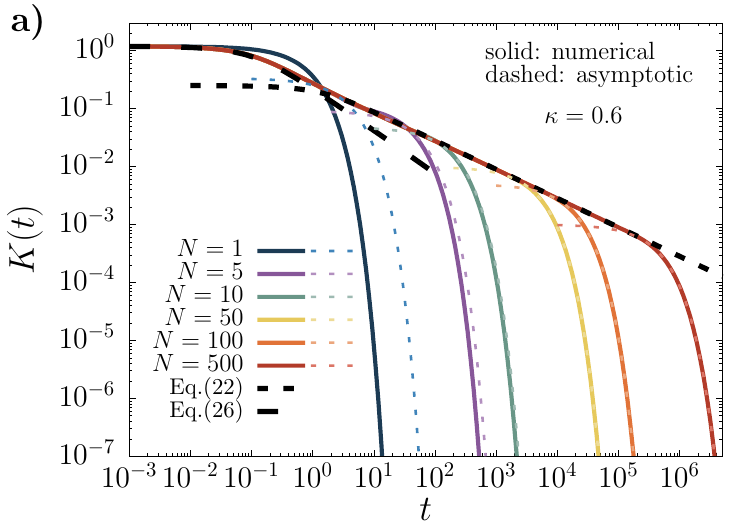}
	\includegraphics[width=\columnwidth]{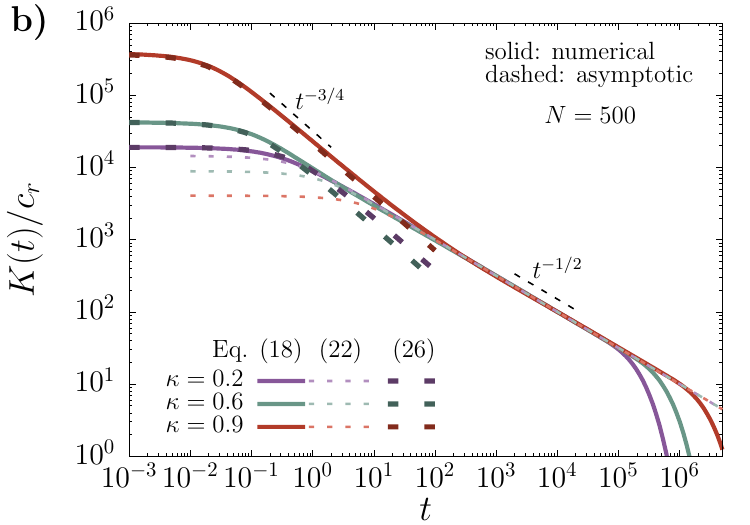}
	\caption{a) Memory kernel $K(t)$ as a function of time $t$ for $\kappa = 0.6$ and for various polymer arm length $N$. The black long dashed line corresponds to the analytical expression Eq.~(\ref{eq:semikernelsmall}) valid in the limit $t\ll\tau_l'$ while the black short dashed line corresponds to Eq.~(\ref{eq:semikernel1intermediate}) valid in the limit $t\gg\tau_l'$. The solid lines correspond to the numerical evaluation of the Kernel expression Eq.~(\ref{eq:kernel}). The respective colored dashed lines are the long time limit $t\gg \tau_R$ given in Eq.~(\ref{eq:semikernel1large}). b) Memory kernel $K(t)$ as a function of time $t$ for a polymer length $N=500$ for three values of $\kappa = (0.2,0.6,0.9)$. For readability, the memory kernel is rescaled by a constant $c_r$ given below Eq.~(\ref{eq:semiflexiblekernelsummary}). The solid lines represent the numerical evaluation of the kernel Eq.~(\ref{eq:kernel}), the thick dashed lines correspond to the initial behavior Eq.~(\ref{eq:semikernelsmall})  for $t\ll \tau_l'$, while the thin dashed lines represent the behavior Eq.~(\ref{eq:semikernel1intermediate}) for $t\gg\tau_l'$. Both the spring constant $k$ and friction coefficient $\gamma$ are set to unity.}
	\label{fig:Fig3}
\end{figure}
For a large number of beads $N\gg1$, we can approximate the summation by an integral. The kernel $K_1(t)$ in Eq.~(\ref{eq:semikernel1}) governs the behavior at sufficiently large times, enabling us to simplify the argument within the exponential by considering only the lowest order of modes. Its expression reads
\begin{eqnarray}\nonumber
K_1(t) &\simeq& 8k(\lambda - \mu) \int_0^1 {\rm d}p \; {\rm cos}^2\left( \frac{p\pi}{2}\right){\rm e}^{-\frac{8(\lambda -\mu)kt}{\gamma} {\rm sin}^2\left( \frac{p\pi}{2}\right)} \\  \label{eq:semikernel1intermediate}
&=&2k'{\rm e}^{-\frac{2k't}{\gamma}}\left[I_0\left(\frac{2k't}{\gamma}\right)+I_1\left(\frac{2k't}{\gamma}\right)\right]
\end{eqnarray}
where $k' = 2(\lambda - \mu)k$. Identifying the time scale 
\begin{equation}
\tau_l' = \frac{\gamma}{4(\lambda - \mu)k}=\frac{\gamma}{2k'},
\end{equation}
the kernel $K_1(t)$ behaves as a constant for short time $t\ll \tau_l'$ and behaves as a power-law $t^{-1/2}$ for long time $t\gg \tau_l$. 
For the kernel $K_2(t)$ in Eq.~(\ref{eq:semikernel2}), it is relevant for sufficiently short times, allowing us to simplify the argument within the exponential by considering only the highest order of modes. Its expression is given by:
 \begin{eqnarray}\nonumber
K_2(t) &\simeq& 4\mu k\int_0^1 {\rm d}p \; {\rm sin}^2\left( p\pi \right){\rm e}^{-\frac{16 \mu kt}{\gamma} {\rm sin}^4\left( \frac{p\pi}{2}\right)} \\  \label{eq:semikernel2intermediate}
&=& 2k\mu \;_2F_2 \left(\frac{3}{4},\frac{5}{4};\frac{3}{2},2;-\frac{16\mu k t}{\gamma}\right)
\end{eqnarray}
where $_2F_2(a,b;c,d;z)$ is the hypergeometric function. We identify the characteristic time scale 
\begin{equation}
\tau_s = \frac{\gamma}{16\mu k}
\end{equation}
which separates the initial constant behavior of the kernel $K_2(t)$ for small times $t\ll\tau_s$ and the final power-law behavior $t^{-3/4}$ for large times $t\gg\tau_s$. Because of the relation between the Lagrangian multiplier $\lambda$ and $\mu$ given in Eq.~(\ref{eq:lagrangemultipliers}), we can calculate the ratio between the two timescales $\tau_s$ and $\tau_l'$ and obtain $\tau_s/\tau_l' = (1-\kappa)^2/8\kappa$. This ratio is always less than unity, ensuring that the two timescales are consistently ordered in the same manner, with $\tau_s < \tau_l'$. 

For a short time $t\ll\tau_l$, the kernel $K(t)$ behaves according to $K(t)\simeq K_2(t) \left[1+K_1(t\rightarrow0)/K_2(t\rightarrow0)\right]$ which reads
\begin{eqnarray}\label{eq:semikernelsmall}
K(t) = 2k(2\lambda-\mu)\;_2F_2 \left(\frac{3}{4},\frac{5}{4};\frac{3}{2},2;-\frac{16\mu k t}{\gamma}\right).
\end{eqnarray}
Then for a time $t\gg \tau_l$, the contribution $K_2(t)$ becomes irrelevant such that 
\begin{eqnarray} \label{eq:semikernelintermediate}
K(t)\simeq K_1(t)=2k'{\rm e}^{-\frac{2k't}{\gamma}}\left[I_0\left(\frac{2k't}{\gamma}\right)+I_1\left(\frac{2k't}{\gamma}\right)\right].
\end{eqnarray}

For the asymptotic behavior of the kernel $K(t)$, the mode $p=1$ plays a dominant role in the sum. Assuming that the number of beads $N$ is sufficiently large, we can approximate trigonometric functions using the first order of their Taylor expansion. This approximation yields
\begin{eqnarray} \label{eq:semikernel1large}
K(t) = \frac{16(\lambda-\mu)  k}{2N+1}  e^{-\frac{2(\lambda-\mu)kt}{\gamma}\frac{\pi^2}{(2N+1)^2}}
\end{eqnarray}
where we identify the final time scale
\begin{equation}
\tau_R' = \frac{\gamma(2N+1)^2}{2(\lambda -\mu)\pi^2k}=\frac{\gamma(2N+1)^2}{\pi^2k'}.
\end{equation}
As shown in Fig.~\ref{fig:Fig3}a, the asymptotic solution Eq.~(\ref{eq:semikernel1large}) provides a proper description of the kernel for sufficiently long polymers $N>50$. Indeed, the colored dashed lines are in agreement when $t\gg \tau_R$ with the solid lines which correspond to the numerical evaluation of the kernel Eq.~(\ref{eq:kernel}). Fig.~\ref{fig:Fig3}b exhibits the different behaviors of the kernel over the entire time range and the perfect agreement between the numerical evaluation of the kernel Eq.~(\ref{eq:kernel}) (solid lines), the approximation at short times $t\ll \tau_l'$ given in Eq.~(\ref{eq:semikernelsmall}) (thick dashed lines) and the approximation at large times $t\gg\tau_l'$ given in Eq.~(\ref{eq:semikernel1intermediate}) (thin dashed lines). The following set of equations summarizes the behaviors of the kernel
\begin{eqnarray} \label{eq:semiflexiblekernelsummary}
t\ll\tau_s :&\; K(t)= 2k(2\lambda -\mu)\\ \nonumber
\tau_s \ll t \ll \tau_l' :&\; K(t) = C_{\rm semi} t^{-3/4}\\ \nonumber
\tau_l' \ll t \ll \tau_R' :&\; K(t) = c_{r}t^{-1/2} \\ \nonumber
t\gg\tau_R' :&\; K(t) = \frac{16 k(\lambda-\mu)}{2N+1} e^{-\frac{t }{\tau_R'}},
\end{eqnarray}
where the prefactor appearing in the Rouse regime $\tau_l' \ll t \ll \tau_R'$ is $c_r= 2k'\sqrt{\frac{2\tau_l'}{\pi}}$ and where the derivation of $C_{\rm semi}$ is given in appendix C: $C_{\rm semi} = 2k(2\lambda-\mu)\frac{\Gamma(2)B\left(\frac{1}{2},\frac{3}{4}\right)}{\Gamma(3/4)}\left(\frac{2}{\pi}\right)^{1/2} \left(\frac{\gamma}{16\mu k}\right)^{3/4}$. 

\subsection{Kernel in the stiff-polymer limit}
In this section, we study the behavior of the kernel $K(t)$ given in Eq.~(\ref{eq:kernel}) when the polymer is stiff. In such a case, the parameter $\kappa$ tends to unity $\kappa\rightarrow 1$. Thus, the parameter $\lambda - \mu$ becomes negligible such that the term $(\lambda-\mu)+2\mu {a'_p}^2 $ is dominated by $2\mu {a'_p}^2$. It can be rewritten using Eq.~\eqref{eq:lagrangemultipliers} as
\begin{eqnarray}
\frac{1-\kappa}{1+\kappa}\ll \frac{4\kappa}{1-\kappa^2}{\rm sin}^2\left(\frac{(2p-1)\pi}{2(2N+1)}\right).
\end{eqnarray}
\begin{figure}[b]
\centering
\includegraphics[width=\columnwidth]{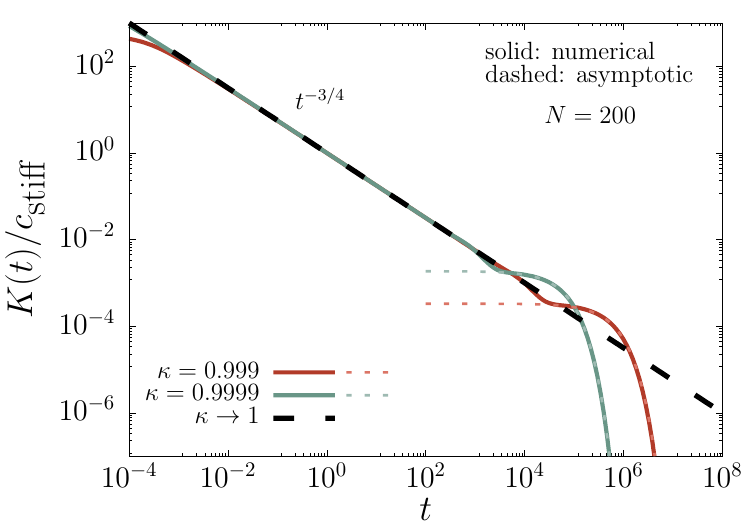}
\caption{Memory kernel $K(t)$ as a function of time $t$ for various bending stiffness $\kappa$. The memory kernel $K(t)$ has been scaled by the prefactor $c_{\rm stiff}$ defined in Eq.~\eqref{eq:stifflimitkernel}. The black dashed line corresponds to the time dependence $t^{-3/4}$ of Eq.~\eqref{eq:stifflimitkernel}. The colored dashed line represents the large time behavior of the kernel given in Eq.~\eqref{eq:stiffkernellarge}.}
\label{fig:Fig4}
\end{figure}
If this inequality holds for mode $p=1$, it remains true for all modes. Furthermore, assuming a large number of beads $2N+1 \gg 1$ the sine function can be approximated by its argument $\mathrm{sin} (x) \approx x$. Consequently, the inequality simplifies to $(2N+1)^2(1-\kappa)^2\ll\kappa\pi^2$ or equivalently $l^2\ll\kappa\pi^2$. Thus, it becomes evident that in the limit $\kappa \rightarrow 1$,  the persistence length $L_p$ must be much larger than the contour length $L$. In such cases, the kernel can be expressed as
\begin{eqnarray} \label{eq:stiffkernel}
K(t) = \frac{8\mu k}{2N+1}\sum_{p=1}^N& a_p^2 e^{-\frac{16\mu kt}{\gamma}{a'_p}^4 }.
\end{eqnarray}
Considering the number of beads to be large $N\gg 1$, we can approximate the sum using an integral, where we make the substitution $\frac{p}{2N+1}\rightarrow p$ as the integrating variable. This allows us to express it as
\begin{eqnarray}
K(t) &\simeq& 4\mu k\int_0^1 {\rm d}p \; {\rm sin}^2\left( p\pi \right){\rm e}^{-\frac{16 \mu kt}{\gamma} {\rm sin}^4\left( \frac{p\pi}{2}\right)} \\  
&=& 2k\mu \;_2F_2 \left(\frac{3}{4},\frac{5}{4};\frac{3}{2},2;-\frac{16\mu k t}{\gamma}\right).
\end{eqnarray}
In the limit of $\kappa \rightarrow 1$, or equivalently for $\mu \gg 1$, this expression can be further reduced to 
\begin{eqnarray}\label{eq:stifflimitkernel}
K(t) &\simeq& (k\mu)^{1/4} \gamma^{3/4}\frac{\Gamma(2)B\left(\frac{1}{2},\frac{3}{4}\right)}{4\Gamma\left(\frac{3}{4}\right)}\left(\frac{2}{\pi}\right)^{1/2} t^{-3/4} \nonumber \\
&=& c_{\rm stiff} \; t^{-3/4} 
\end{eqnarray}
where we trivially read off the time dependence and where $c_{\rm stiff}$ is a prefactor that encompasses the details of the model. As anticipated, we retrieve the power-law behavior of the kernel, a characteristic of the semiflexible polymer, namely $t^{-3/4}$, as depicted in Fig.~\ref{fig:Fig4}. Thus, for a stiff polymer, the mean-squared displacement (MSD) will exhibit a behavior following $t^{3/4}$. Finally,  it is worth noting that the large-time behavior can be straightforwardly determined by examining the mode $p=1$ of the kernel \eqref{eq:stiffkernel}
\begin{eqnarray} \label{eq:stiffkernellarge}
K(t) = \frac{2\pi^2\mu k}{(2N+1)^3}  e^{-t/\tau_W}
\end{eqnarray}
where $\tau_W = \frac{\gamma(2N+1)^4}{\pi^4\mu k}$.

\section{Dynamics of the GLE}
In this section, we analyze the dynamics of the resulting GLE~(\ref{eq:GLE}). We derive the mean-square displacement (MSD), which reads 
\begin{eqnarray}\label{eq:MSD} 
   \langle (r(t)-r(0))^2 \rangle = 6k_BT{\mathcal L}^{-1}\left[s^{-2}(\gamma+\tilde{K}(s))^{-1}\right](t)
\end{eqnarray}
where ${\mathcal L}^{-1}$ is the inverse Laplace transform and $\tilde{K}(s)$ is the Laplace transform of the kernel $K(t)$. 

First, we briefly discuss the limit of the Rouse chain with $\kappa =0$. The asymptotic behavior in each time domain reads 
\begin{eqnarray} \label{eq:rouseMSDsummary}
t\ll\tau_l :&\;  {\rm MSD}(t)\simeq \frac{2k}{\gamma}t \\ \nonumber
\tau_l \ll t \ll \tau_R :&\;  {\rm MSD}(t)\simeq 2\sqrt{\frac{k}{\pi\gamma}} \sqrt{t} \\ \nonumber
t\gg\tau_R :&\;\;  {\rm MSD}(t) \simeq \frac{2k}{2N+1}\left[\frac{8 \gamma}{\pi^2 {\rm e}}+\frac{4\gamma}{\pi^{3/2}}\right]^{-1}t
\end{eqnarray}
which is in agreement with previous studies \cite{Vand2017a,Joo2020, Tian2022, Teje2023}. The details of the derivation of the MSD are given in Appendix B. In Fig.~\ref{fig:Fig5}a, we plot the MSD from the formal expression Eq.~(\ref{eq:MSD}) evaluated numerically using the Gaver-Stehfest algorithm \cite{Steh1970} together with the limiting behaviors from Eq.~(\ref{eq:rouseMSDsummary}) and the previously derived analytical result \cite{Joo2020}. The agreement is perfect except in the region $t\ll\tau_l$ because of the domain of validity of the solution from the analytic expression in Ref.~\cite{Joo2020}. The behavior of the MSD can be interpreted as the following: For $t\ll\tau_l$, the equation of motion reduces to $\gamma \dot{r} = -2k r(t) + \xi(t)$ such that the non-Markovian term behaves as a harmonic potential. The particle does not feel the entirety of the polymers and only the nearest neighbors create this effective potential. Therefore, the MSD is Fickian $\langle (r(t)- r(0))^2\rangle \sim t$. For large times $t\gg \tau_R$, as the characteristic time scale $\tau_R$ scales as $N^2$, the exponential term dominates the dynamics and also leads to a Fickian behavior and is understood as the dynamics of the center of mass. In the intermediate regime, the collective dynamics dominates and gives the Rouse regime $\langle (r(t)- r(0))^2\rangle \sim t^{1/2}$. 

For the semiflexible polymer, we have also derived the limiting behaviors from our GLE (see Appendix B for details), and they read
\begin{eqnarray} \label{eq:semiflexibleMSDsummary}
t\ll\tau_s :&\;  {\rm MSD}(t)\simeq \frac{2k}{\gamma}t \\ \nonumber
\tau_s \ll t \ll \tau_l' :&\;  {\rm MSD}(t)\simeq \frac{2k}{\Gamma(1/4)C_{\rm semi}}t^{3/4} \\ \nonumber
\tau_l' \ll t \ll \tau_R' :&\;  {\rm MSD}(t)\simeq 2\sqrt{\frac{k}{\pi\gamma}\frac{1+\kappa}{1-\kappa}} \sqrt{t} \\ \nonumber
t\gg\tau_R' :&\;\;  {\rm MSD}(t) \simeq \frac{2k}{2N+1}\left[\frac{8 \gamma}{\pi^2 {\rm e}}+\frac{4\gamma}{\pi^{3/2}}\right]^{-1}t
\end{eqnarray}
where we recall the four distinct timescales
\begin{eqnarray} \label{eq:semiflexibletimescales}
\tau_s &=&= \frac{\gamma}{16 k} \frac{1-\kappa^2}{\kappa}\\ \nonumber
\tau_l' &=& \frac{\gamma}{2k}\frac{1+\kappa}{1-\kappa}\\ \nonumber
\tau_R' &=&  \frac{\gamma(2N+1)^2}{\pi^2k}\frac{1+\kappa}{1-\kappa} \\ \nonumber
\tau_W &=& \frac{\gamma(2N+1)^4}{\pi^4 k} \frac{1-\kappa^2}{\kappa}.
\end{eqnarray}
\begin{figure}[t]
\centering
\includegraphics[width=\columnwidth]{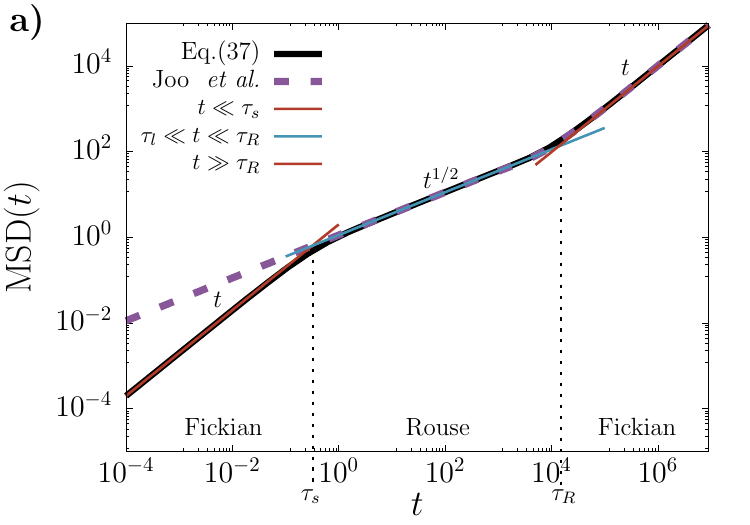}
\includegraphics[width=\columnwidth]{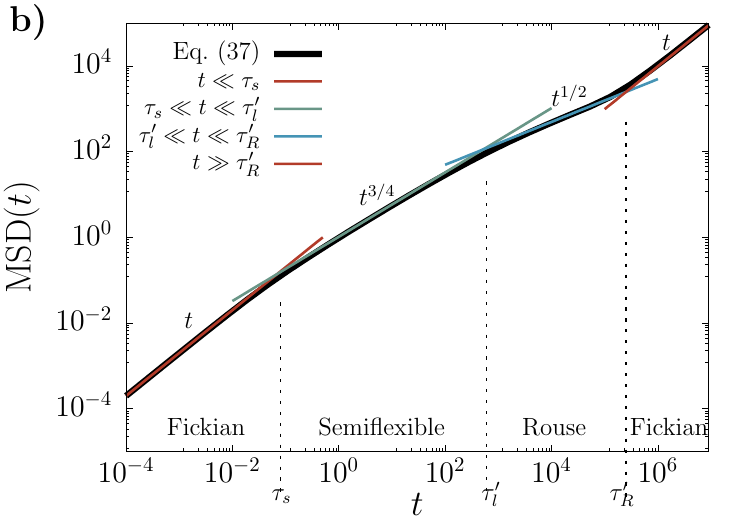}
\caption{Mean-squared displacement: the thick black lines represent the numerical evaluation of the MSD Eq.~(\ref{eq:MSD}). a) Rouse case $\kappa =0$, the thick purple dotted line is the MSD obtained in \cite{Joo2020}, the thin colored straight lines represent the limiting behavior of Eq.~(\ref{eq:rouseMSDsummary}). b) Semiflexible case $\kappa =0.9$, the thin colored straight lines represent the limiting behavior of Eq.~(\ref{eq:semiflexibleMSDsummary}). The other parameters are $N=100$, $k=1$ and $\gamma =1$.}
	\label{fig:Fig5}
\end{figure}
For a semiflexible polymer, the subdiffusive regime starts earlier than the Rouse behavior $\tau_s<\tau_l'$. As the bending rigidity increases, the smallest timescale $\tau_s$ decreases because the stiffness imposes more restriction on the displacement of beads, or said differently, because the stiffness creates an additional constraint on the harmonic interaction between the nearest beads. The subdiffusive regime follows a power-law $t^{3/4}$, characteristic of the semiflexible polymer, and lasts until the time $\tau_l'$ which increases with the bending rigidity $\kappa$. As $\kappa$ increases, even though the beads start from a lower MSD, they have to reach the same Fickian regime after the largest timescale, therefore it requires the subdiffusive regime to have a higher slope than in the Rouse regime. After $\tau_l'$, the MSD of the central bead exhibits a crossover from subdiffusive to the Rouse behavior as a function of the bending rigidity parameter $\kappa$ (Fig.~\ref{fig:Fig5}b). Remarkably, in the Rouse regime $\tau_l \ll t \ll \tau_R$ in Eq.~(\ref{eq:rouseMSDsummary}) and $\tau_l' \ll t \ll \tau_R'$ in Eq.~(\ref{eq:semiflexibleMSDsummary}), the beads in the semiflexible polymer always move faster than their counterparts in the Rouse chain because the prefactor $2\sqrt{\frac{k}{\pi\gamma}\frac{1+\kappa}{1-\kappa}}$ in Eq.~(\ref{eq:semiflexibleMSDsummary}) is always larger than $2\sqrt{\frac{k}{\pi\gamma}}$ in Eq.~(\ref{eq:rouseMSDsummary}).  Finally, in the limit of rigid rod $\kappa \rightarrow 1$, this crossover disappears and only the subdiffusive behavior remains: because the relaxation time $\tau_W$ decreases with $\kappa$, there exists a specific value $\kappa*$ such that the relaxation time $\tau_W$ is about the crossover time $\tau_l'$. For any bending rigidity $\kappa$ such $\kappa^*\leq\kappa<1$, the Rouse regime disappears and the polymer enters the stiff polymer limit and only displays the subdiffusive behavior $t^{3/4}$. All these observations are in qualitative agreement with previous work \cite{Teje2023} that exactly solves the MSD of the Winkler semiflexible polymer.

\section{\label{sec:conclusion}Discussion}


In this work, starting from the bead-spring model of the Gaussian semiflexible polymer  \cite{Wink1994}, we derived the Generalized Langevin Equation satisfied by a tagged monomer. We have provided a thorough analysis of the memory kernel $K(t)$ depending on the bending rigidity of the polymer. This allowed us to obtain a complete and accurate description of the polymer dynamics at different time scales and understand how the bending rigidity affects the non-Markovian behavior. For the Rouse case $\kappa=0$, we reproduce the known results \cite{Liza2010, Panja2010,Vand2017a}. For the semiflexible case $\kappa>0$, even though our derivation requires an approximation, our resulting GLE~\eqref{eq:GLE} together with the memory kernel in Eq.~(\ref{eq:kernel})  and the effective noise in Eq.~(\ref{eq:effectivenoise}) exhibits a rich behavior. 

Between the Fickian dynamics $t$ at early and large times, we observe a crossover from semiflexible $t^{3/4}$ to flexible $t^{1/2}$ behavior. Such a crossover has been observed and studied in the bead-spring worm-like chain (WLC), for the center of mass dynamics \cite{Stein2008} and for the end monomer \cite{Hinc2009}, it appears for small persistence length. Therefore, even though the Gaussian semiflexible polymer is, in a sense, simpler than the WLC because it allows Gaussian fluctuations of the length, this crossover is observed which means that its origin is not from the longitudinal fluctuations. 

For a sufficiently large bending stiffness $\kappa$ such that  $\tau_W\ll\tau'_R$, the crossover disappears. For an infinitely long polymer with $N\rightarrow \infty$, the kernel reduces to a pure power-law $t^{-3/4}$ and the noise satisfies the fluctuation-dissipation theorem. For times larger than the smaller time scale $t\gg\tau_s$, one can neglect the terms $\gamma \mathbf{\dot{r}}$ and $\xi(t)$ in Eq.~\eqref{eq:GLE} and the equation reduces to the fractional Langevin equation (FLE)
\begin{eqnarray}
    \int_0^t {\rm d} \tau \;K(t-\tau)\mathbf{\dot{r}}(\tau) = \Gamma(1/4)\frac{{\rm d}^{3/4}}{{\rm d}t^{3/4}}\mathbf{r}(t) = \Xi(t)
\end{eqnarray}
where $\frac{{\rm d}^{3/4}}{{\rm d}t^{3/4}}\mathbf{r}(t)$ is the Caputo fractional derivative of order $3/4$. The thermal noise $\Xi(t)$ whose auto-covariance is a power-law $\langle \Xi(t)\Xi(t') \rangle \sim \|t-t'\|^{2H-2}$, is realized by the fractional Gaussian noise of Hurst exponent $H=5/8$. As is known, this type of fractional Langevin equation describes viscoelastic subdiffusion \cite{Jeon2010,Zhu2023, Lutz2001,Kurs2013, Buro2008}. We showed that, in the context of polymer, this FLE is only relevant in the stiff polymer limit when the kernel displays a single power-law. For other values of the bending rigidity, the kernel will display two power-laws which, to our current knowledge, have not been systematically studied.

Our derivation can be easily extended to the case of a semiflexible polymer with an additive force  $\eta(t)$ acting only on the tagged monomer. Assuming that the active force is sufficiently weak so that our approximation remains valid, the additive force is unchanged by the procedure, and the final equation is our result in Eq.~(\ref{eq:GLE}) to which we simply add the extra force. This framework can be used directly for the case of a bead-spring model with an active force $F_{\rm ext}(t)$ acting on the tagged central bead. In such a case, the equation will be
\begin{eqnarray}
\gamma \dot{\mathbf r} = -\int_0^t d\tau \, K(t-\tau) \dot{\mathbf r}(\tau) + \Xi(t)+ \xi(t) + F_{\rm ext}(t).
\end{eqnarray}
It constitutes a basic framework that can be used in numerical or analytical studies, e.g., to obtain the position and velocity autocorrelation functions, mean-square displacement or to analyze the ergodicity of such systems \cite{Joo2020,Han2023,Joo2023}. 

In conclusion, our work provides a detailed derivation of the GLE satisfied by a tagged monomer as well as a comprehensive analysis of the motion of the central bead through the MSD in both finite Rouse polymer and finite Gaussian semiflexible polymer models over the entire timescale. The derived GLE is an adequate framework for further studies on polymer dynamics and it is our hope that this work will inspire further research in this field.



\section*{Acknowledgements}
We acknowledge the support from the National Research Foundation (NRF) of Korea, No.~RS-2023-00245195 (X.D.), No.~RS-2023-00218927 \& No.~RS-2024-00343900 (J.-H.J.).

\section*{Author declarations}
\subsection*{Conflict of Interest}
The authors have no conflicts to disclose.
\subsection*{Author Contributions}
\textbf{Xavier Durang:} Conceptualization (equal); Formal analysis (lead); Investigation (lead); Writing – original draft (lead); Writing – review \& editing (equal). 
\textbf{Jae-Hyung Jeon:} Conceptualization (equal); Supervision (lead); Writing – review \& editing (equal).

\section*{Data Availability}
The data supporting this study's findings are available upon request.

\appendix
\section{Potential energy}
We rewrite the potential energy of the left arm and right arm using the vectors of position coordinates ${\mathbf{R}^L}=(\mathbf{r}_0, \mathbf{r}_1, ..., \mathbf{r}_N)$ and ${\mathbf{R}^R}=(\mathbf{r}_{N+2}, \mathbf{r}_{N+3}, ..., \mathbf{r}_{2N+1})$ such that
\begin{equation}\label{eq:Uch_appendix}
U_{\rm ch} = \frac{k}{2}\left(\mathbf{R}^L H^L {\mathbf{R}^L}^T +\mathbf{R}^R H^R {\mathbf{R}^R}^T \right)
\end{equation}
where $H^{L,R}$ are coupling matrices. Their expressions are given by 
\begin{equation}
H^L=
\begin{bmatrix}
A_1  & B_1  & C & 0 & \cdots & \cdots & 0 \\
B_1  & A_{12} & B  & \ddots  && & \vdots \\
C & B  & A & \ddots   & &  & \vdots \\
\vdots & \ddots & \ddots  & \ddots & \ddots & \ddots & \vdots \\
\vdots  & & & \ddots & A &  B  & C \\
\vdots  & & & \ddots & B  & A &  B  \\
0 & \cdots &  \cdots & 0 & C  & B  & A  \\
\end{bmatrix}
\end{equation}
for the left arm and 
\begin{equation}
H^R=
\begin{bmatrix}
A  & B  & C & 0 & \cdots & \cdots & 0 \\
B  & A & B  & \ddots  && & \vdots \\
C & B  & A & \ddots   & &  & \vdots \\
\vdots & \ddots & \ddots  & \ddots & \ddots & \ddots & \vdots \\
\vdots  & & & \ddots & A &  B  & C \\
\vdots  & & & \ddots & B  & A_{12} &  B_1  \\
0 & \cdots &  \cdots & 0 & C  & B_1  & A_1  \\
\end{bmatrix}
\end{equation}
 for the right arm we have. 
The parameters appearing in the coupling matrices are the following
\begin{eqnarray}
A_1&=& \lambda_1 \nonumber \\
A_{12}&=& \lambda + \lambda_1 +\mu \nonumber \\
A&=& 2\lambda +\mu \nonumber \\
B_1&=& -\lambda_1 -\mu/2 \nonumber\\
B&=& -\lambda -\mu \nonumber\\
C&=& \mu/2. \nonumber
\end{eqnarray}

\section{MSD}
To calculate the MSD, we use the Laplace transform $\tilde{f}(s) = \int_0^\infty {\rm d}t {\rm e}^{-st}f(t)$. Under this transformation, the GLE Eq.~(\ref{eq:Langevinposition}) reads
\begin{eqnarray}\label{eq:MSD_appendix}
    \left(\gamma + \tilde{K}(s)\right) (s\,\mathbf{r}(s)-s_0)= \tilde{\xi}(s) + \tilde{\Xi}(s)
\end{eqnarray}
which leads to 
\begin{eqnarray}
    \tilde{\mathbf{r}}(s)= \frac{\mathbf{r}_0}{s} + \frac{\tilde{\Xi}(s)}{s\left(\gamma + \tilde{K}(s)\right)}.
\end{eqnarray}
Using the fluctuation-dissipation relation for both noises, the expression of the correlation function $\langle \tilde{r}(s) \tilde{r}(s') \rangle$ follows straightforwardly
\begin{eqnarray}
    & &\langle \tilde{\mathbf{r}}(s) \tilde{\mathbf{r}}(s') \rangle = \frac{\langle \mathbf{r}_0^2 \rangle}{s s'}\\ \nonumber
    & &+ \frac{3k_BT}{s+s'}\left(\frac{1}{ss'\left(\gamma + \tilde{K}(s)\right)}+\frac{1}{s's\left(\gamma + \tilde{K}(s')\right)}\right).
\end{eqnarray}
This expression can be recast as 
\begin{eqnarray}
    \langle \tilde{\mathbf{r}}(s) \tilde{\mathbf{r}}(s') \rangle &=& \frac{\langle \mathbf{r}_0^2 \rangle}{s s'} \\ \nonumber
    & &+ \frac{3k_BT}{s+s'}\left(\frac{\tilde{F}(s)}{s}+\frac{\tilde{F}(s')}{s'} + \frac{\tilde{F}(s)+\tilde{F}(s')}{s+s'}\right)
\end{eqnarray}
where $\tilde{F}(s)= \frac{1}{s^2\left(\gamma + \tilde{K}(s)\right)}$. Its inverse Laplace transform reads
\begin{eqnarray} \label{eq:appendix:C5}
    \langle \mathbf{r}(t) \mathbf{r}(t') \rangle = \langle \mathbf{r}_0^2 \rangle +3k_BT\left(F(t)+F(t') - F(|t-t'|) \right)
\end{eqnarray}
where $F(t)$ is the inverse Laplace transform of $\tilde{F}(s)$. Finally, the MSD can be calculated using the correlation function given in Eq.~(\ref{eq:appendix:C5})
\begin{eqnarray}
    \langle (\mathbf{r}(t+\tau)-\mathbf{r}(\tau))^2 \rangle &=& \langle \mathbf{r}(t+\tau)^2 \rangle + \langle \mathbf{r}(\tau)^2 \rangle
    \\ \nonumber 
    & & - 2 \langle \mathbf{r}(t+\tau)\mathbf{r}(\tau) \rangle \\ \nonumber 
    &=&6k_BTF(t).
\end{eqnarray}
Thus we showed that the calculation of the MSD reduces to the evaluation of the inverse Laplace transform of $\tilde{F}(s)= \frac{1}{s^2\left(\gamma + \tilde{K}(s)\right)}$. Using the expression of the kernel Eq.~(\ref{eq:kernel}), the function $\tilde{F}(s)$ reads
\begin{eqnarray}
    \tilde{F}(s)= \frac{1}{\gamma +\frac{2ks^2}{2N+1}\sum_{p=1}^N \frac{\frac{a_p^2}{{a'_p}^2}\left(2(\lambda-\mu)+4\mu {a'_p}^2\right) }{s+\frac{4k}{\gamma}{a'_p}^2\left(2(\lambda-\mu)+4\mu {a'_p}^2\right)}}
\end{eqnarray}

The MSD can then be calculated by numerically evaluating the inverse Laplace transform with the Gaver-Stehfest algorithm \cite{Steh1970}. To go further and obtain some analytical results, we approximate the kernel by its limiting behavior in each time domain.
\begin{figure}[!tb]
\centering
	\includegraphics[width=\columnwidth]{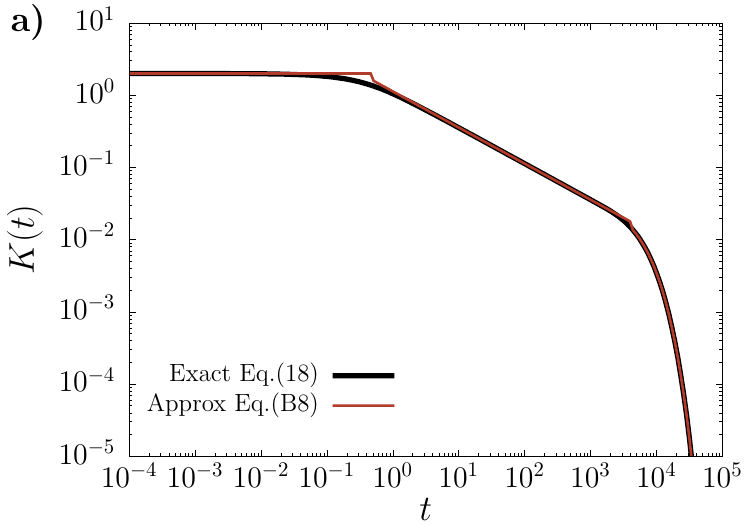}
	\includegraphics[width=\columnwidth]{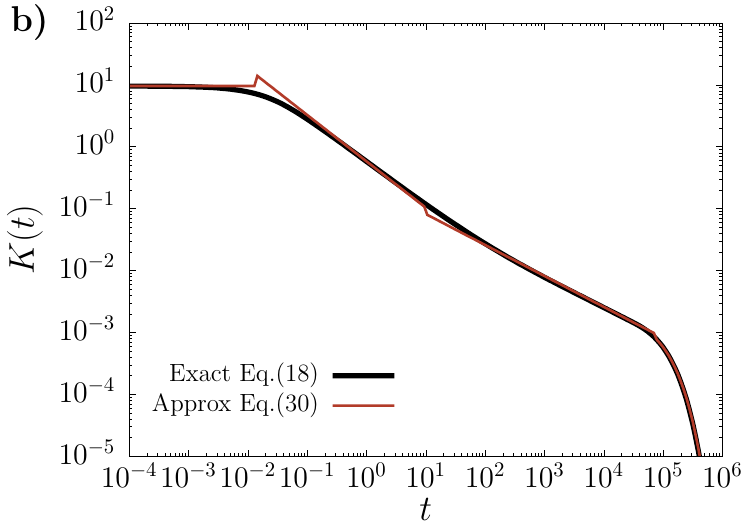}
	\caption{Approximation of the kernel. The thick black line is the numerical evaluation of the exact expression of the kernel Eq.~(\ref{eq:kernel}): a) for the Rouse case $\mu = 0$, b) for the semiflexible polymer $\mu = 0.9$.  The thinner red line is the approximation we use: it is the piecewise function given by the inverse Laplace of\ Eq.~(\ref{eq:rousekernelsummaryappendix}) for the Rouse case and in Eq.~(\ref{eq:semiflexiblekernelsummary}) for the semiflexible case. The other parameters are $N=100$, $k=1$ and $\gamma =1$.}
	\label{fig:FigS1}
\end{figure}
For the Rouse polymer, as depicted in Fig.~(\ref{fig:FigS1}a), the behavior of the kernel in each time domain is 
\begin{eqnarray}\label{eq:rousekernelsummaryappendix}
t\ll\tau_l &\; K(t)= 2k\\ \nonumber
\tau_l \ll t \ll \tau_R &\; K(t) = 2k\sqrt{\frac{2\tau_l}{\pi t}} \\ \nonumber
t\gg\tau_R &\; K(t) = \frac{8 k}{2N+1} e^{-\frac{t }{\tau_R}}.
\end{eqnarray}
and the approximate Laplace transform of the kernel reads
\begin{eqnarray} \nonumber
    \tilde{K}(s) &\simeq& \int_0^{\tau_l} {\rm d}t \,2k \,{\rm e}^{-st} + \int_{\tau_l}^{\tau_R} {\rm d}t \sqrt{\frac{8k^2\tau_l}{\pi t}}{\rm e}^{-st} 
    \nonumber 
    \\ & & + \int_{\tau_R}^{\infty} {\rm d}t \frac{8 k}{2N+1} {\rm e}^{-\frac{t }{\tau_R}}{\rm e}^{-st} .
\end{eqnarray}
Plugging it in the function $\tilde{F}(s)$, we obtain
\begin{eqnarray}
    \tilde{F}(s) &=& s^{-2}\left[\gamma + \frac{2k}{s}\left(1-{\rm e}^{-s\tau_l}\right) +\frac{8k\tau_R{\rm e}^{-\tau_Rs-1}}{(2N+1)(\tau_Rs+1)}\right.
    \nonumber
    \\
    & & \left.+ 2\sqrt{\frac{k\gamma}{s}}\left({\rm erf}\sqrt{\tau_R s}-{\rm erf}\sqrt{\tau_l s}\right)\right]^{-1}.
\end{eqnarray}
There are three time domains and for each, we give the approximate behavior
\begin{eqnarray} \label{eq:rousekernellaplacesummaryappendix}
s\gg (1/\tau_l,1/\tau_R) :&\; \tilde{F}(s)\simeq \gamma^{-1} s^{-2}\\ \nonumber
1/\tau_R\gg s \gg 1/\tau_l :&\; \tilde{F}(s)\simeq \sqrt{\frac{k}{\gamma}}s^{-3/2} \\ \nonumber
(1/\tau_l,1/\tau_R) \gg s :&\;\; \tilde{F}(s) \simeq \frac{s^{-2}}{2N+1}\left[\frac{8 \gamma}{\pi^2 {\rm e}}+\frac{4\gamma}{\pi^{3/2}}\right]^{-1}.
\end{eqnarray}
This approximation will only be valid if the characteristic time scales are well separated $\tau_R\gg\tau_l$. The MSD follows from the previous equation Eq.~(\ref{eq:rousekernellaplacesummaryappendix}) by taking the inverse Laplace transform and we obtain
\begin{eqnarray} \label{eq:rouseMSDsummaryappendix}
t\ll\tau_l :&\;  {\rm MSD}(t)\simeq \frac{2k}{\gamma}t \\ \nonumber
\tau_l \ll t \ll \tau_R :&\;  {\rm MSD}(t)\simeq 2\sqrt{\frac{k}{\pi\gamma}} \sqrt{t} \\ \nonumber
t\gg\tau_R :&\;\;  {\rm MSD}(t) \simeq \frac{2k}{2N+1}\left[\frac{8 \gamma}{\pi^2 {\rm e}}+\frac{4\gamma}{\pi^{3/2}}\right]^{-1}t.
\end{eqnarray}

The procedure extends straightforwardly for the case of the semiflexible polymer. We approximate the kernel by its piecewise function Eq.~(\ref{eq:semiflexiblekernelsummary}), see Fig.~(\ref{fig:FigS1}b), we calculate its Laplace transform and then the inverse Laplace transform of the function $\tilde{F}(s)$. We obtain
\begin{eqnarray}
t\ll\tau_s :&\;  {\rm MSD}(t)\simeq \frac{2k}{\gamma}t \\ \nonumber
\tau_s \ll t \ll \tau_l' :&\;  {\rm MSD}(t)\simeq \frac{2k}{\Gamma(1/4)C_{\rm semi}}t^{3/4} \\ \nonumber
\tau_l' \ll t \ll \tau_R' :&\;  {\rm MSD}(t)\simeq 2k\sqrt{\frac{1}{\pi\gamma k'}} \sqrt{t} \\ \nonumber
t\gg\tau_R' :&\;\;  {\rm MSD}(t) \simeq \frac{2k}{2N+1}\left[\frac{8 \gamma}{\pi^2 {\rm e}}+\frac{4\gamma}{\pi^{3/2}}\right]^{-1}t.
\end{eqnarray}
with $C_{\rm semi} = 2k(2\lambda-\mu)\frac{\Gamma(2)B\left(\frac{1}{2},\frac{3}{4}\right)}{\Gamma(3/4)}\left(\frac{2}{\pi}\right)^{1/2} \left(\frac{\gamma}{16\mu k}\right)^{3/4}$.

\section{Large time behavior of Eq.~(\ref{eq:semikernel2intermediate})}
The hypergeometric function $_2F_2(a,b;c,d;-z)$ that appears in the expression of the kernel $K_2(t)$ of Eq.~(\ref{eq:semikernel2intermediate}) has a negative argument. To extract the large argument behavior, we need to rewrite this expression. Using the integral representation of the hypergeometric function $_2F_2$ \cite{Pari2005}, it reads
\begin{eqnarray}
_{2}F_{2}\left(a,b;c,d;-z\right)=\frac{\Gamma(d)}{\Gamma(b)\Gamma(d-b)}\\ \nonumber
\int_0^1 t^{b-1}(1-t)^{d-b-1}{\rm e}^{-zt}\; _1F_1(c-a,c;zt).
\end{eqnarray}
For our case, the parameters are $c-a=3/4$ and $c=3/2$, such that we can rewrite the $_1F_1$ hypergeometric function by using the identity
\begin{eqnarray}
_1F_1\left(\frac{3}{4},\frac{3}{2};zt\right) = \Gamma\left(\frac{5}{4}\right){\rm e}^{zt/2} \left(\frac{zt}{4}\right)^{-1/4}I_{-1/4}\left(\frac{zt}{2}\right)
\end{eqnarray}
We insert the previous expression into the integral representation of the function $_2F_2$ and we take the asymptotic expansion of the Bessel function $I_{-1/4}(x)\simeq \frac{{\rm e}^x}{\sqrt{2\pi x}}$. It reads 
\begin{equation}
\begin{aligned}
&_{2}F_{2}\left(\frac{3}{4},\frac{5}{4};\frac{3}{2},2;-z\right)\stackrel{z\gg1}{=} \\
&\frac{\Gamma(2)}{\Gamma(3/4)}\left(\frac{2}{\pi}\right)^{1/2} z^{-3/4} \int_0^1 {\rm d}t (1-t)^{-1/4}t^{-1/2}.
\end{aligned}
\end{equation}
Identifying the integral to a Beta function, we obtain the final 
\begin{eqnarray}
& &_{2}F_{2}\left(\frac{3}{4},\frac{5}{4};\frac{3}{2},2;-z\right)\stackrel{z\gg1}{=} \frac{\Gamma(2)B\left(\frac{1}{2},\frac{3}{4}\right)}{\Gamma(3/4)}\left(\frac{2}{\pi}\right)^{1/2} z^{-3/4}
\end{eqnarray}
which will straightforwardly give gives the equation Eq.~(\ref{eq:stifflimitkernel}) in the main text.

\section*{References}
\bibliography{bibliography} 

\end{document}